\documentclass[twocolumn,showpacs,preprintnumbers,amsmath,amssymb]{revtex4}

\usepackage{graphicx}
\usepackage{dcolumn}
\usepackage{bm}

\begin{document}

\title{Thermopower of gapped bilayer graphene}

\author{Lei Hao and T. K. Lee}
 \address{Institute of Physics, Academia Sinica, NanKang, Taipei 11529, Taiwan}

\date{\today}

\begin{abstract}
We calculate thermopower of clean and impure bilayer graphene
systems. Opening a band gap through the application of an
external electric field is shown to greatly enhance the
thermopower of bilayer graphene, which is more than four times
that of the monolayer graphene and gapless bilayer graphene at
room temperature. The effect of scattering by dilute charged
impurities is discussed in terms of the self-consistent Born
approximation. Temperature dependence of the thermopower is
also analyzed.
\end{abstract}


\maketitle

\section{\label{sec:Introduction}Introduction}

Seebeck coefficient, also called as thermopower, measures the
voltage drop across a material in response to a temperature
drop. The achievement  of large thermopower is a prerequisite
to realistic applications in heat to electric energy
conversion. Thermopower, among other thermoelectric properties,
also complements the conductivity in elucidating mechanisms
dominating the transport processes\cite{hwang09}. Recently,
thermopower of monolayer graphene, a peculiar two dimensional
electronic system characterized by a Dirac like relativistic
dispersion, attracts much attention from both
experimental\cite{zuev09,wei09,checkelsky09} and theoretical
groups\cite{lofwander07,dora07,hwang09,yan09,zhu09}.
Calculations taking the effect of charged impurity scattering
into account could explain the experimental results very well.
The experimentally observed deviation from the Mott
relationship at low carrier density is interpreted in terms of
electron-hole puddle formation\cite{hwang09} and also by mixing
of valence band and conduction band states by impurity
scattering\cite{yan09}.

Bilayer graphene is another interesting material system
displaying many unusual properties. Upon applying an external
voltage, a semiconducting gap is induced in the otherwise zero
gap band
structure\cite{mccann06,min07,castro07,oostinga08,zhang09,mak09}.
The gap being tunable by external potential difference between
the two layers introduces a new degree of freedom to bilayer
graphene. Up to now, there is neither experimental nor
theoretical work on thermopower of bilayer graphene. It is the
purpose of this work to partially fill this gap by
theoretically predicting the behavior of thermopower in bilayer
graphene systems.

It is established that charged impurity scattering is primarily
responsible for the transport behavior observed in monolayer
graphene\cite{nomura07,hwang07,adam07}. For bilayer graphene,
the prediction in terms of charged impurity scattering is shown
to be in qualitative agreement with the experimental result of
the conductivity, and the opening of a gap in biased bilayer
graphene is proposed to further improve the
agreement\cite{xiao09}. In the present work, we theoretically
study the thermopower of gapped bilayer graphene. We treat
charged impurity scattering in terms of the self-consistent
Born approximation (SCBA)\cite{shon98,yan08,yan09,peres06}. To
ensure the applicability of SCBA, we restrict our calculations
to relatively clean systems with low impurity concentrations,
where the localization effect is not
severe\cite{nilsson07,koshino08b,mkhitaryan08}. Thermopower as
a function of carrier concentration is mainly calculated at
room temperature. We also study the temperature dependence of
thermopower.

\section{Model and Method}

We consider a bilayer graphene system composed of two graphene
single layers arranged in the Bernal stacking\cite{guinea06}. We start from a tight
binding model incorporating nearest neighboring intralayer and
interlayer hopping terms. An on-site potential energy difference
between the two layers is included to model the effect of an
external voltage. In the presence of impurity, the Hamiltonian
consists of two parts: $\hat{H}=\hat{H}_0+\hat{H}_{imp}$. Without of
magnetic field or magnetic impurities, the two spin flavors are
degenerate. We ignore the spin degree of freedom here and multiply
the results by two for spin dependent quantities. The free part of
the Hamiltonian is then written as
\begin{equation} \label{h0}
\hat{H}_{0}=\sum\limits_{\mathbf{k}}\psi^{\dagger}_{\mathbf{k}}H_{0}(\mathbf{k})\psi_{\mathbf{k}},
\end{equation}
in which the vector of fermion creation operators is defined as
$\psi^{\dagger}_{\mathbf{k}}=(a^{\dagger}_{1\mathbf{k}},
b^{\dagger}_{1\mathbf{k}},b^{\dagger}_{2\mathbf{k}},a^{\dagger}_{2\mathbf{k}})$.
$a_{\alpha\mathbf{k}}^{\dagger}$ and
$b_{\alpha\mathbf{k}}^{\dagger}$ create $\alpha$ layer states
with wave vector $\mathbf{k}$ on the $A$ and $B$ sublattice,
respectively. Up to nearest neighbor hopping,
$H_{0}(\mathbf{k})$ is written
as\cite{mccann06,guinea06,nilsson08,zhang09b,hao09}
\begin{equation}
H_{0}(\mathbf{k})=\begin{pmatrix} \frac{V}{2} & \phi(\mathbf{k}) &
t_{\perp} & 0 \\ \phi^{\ast}(\mathbf{k}) & \frac{V}{2} &0 & 0 \\
t_{\perp} & 0 & -\frac{V}{2} & \phi^{\ast}(\mathbf{k}) \\ 0 & 0 &
\phi(\mathbf{k}) & -\frac{V}{2} \end{pmatrix}.
\end{equation}
$\phi(\mathbf{k})=-t\sum\limits_{j=1}^3
e^{i\mathbf{k}\cdot\boldsymbol{\delta}_{j}}$ describes the
intralayer nearest neighbor hopping with strength $t$. The
three nearest neighbor vectors are defined as
$\boldsymbol{\delta}_{1}=(\frac{1}{2},\frac{\sqrt{3}}{2})a$,
$\boldsymbol{\delta}_{2}=(\frac{1}{2},-\frac{\sqrt{3}}{2})a$
and $\boldsymbol{\delta}_{3}=(-1,0)a$\cite{nilsson08}, $a$=1.42
\AA\ is the shortest carbon-carbon bond length. $t_{\perp}$ is
the nearest-neighbor interlayer hopping energy. In this work,
we take $t$=3 eV and $t_{\perp}$=0.3 eV. $V$ is the potential
energy difference between the first and second layers induced
by a bias voltage. Since for every attainable carrier density,
it is possible to find a bias voltage to make the potential
difference between the two layers as $V$ (when the gap induced
by $V$ is experimentally reachable), we would not consider the
Coulomb interaction between imbalanced electron densities of
the two layers and also neglect the dependence of $V$ on the
carrier density $n$ in this
work\cite{mccann06,ando06,min07,avetisyan09,koshino09}.

For charged impurities, the impurity scattering part of the
Hamiltonian is written as\cite{shon98,koshino06,yan08,yan09}
\begin{equation} \label{himp}
\hat{H}_{imp}=\sum\limits_{i}V_{i}(\mathbf{r}_i)n_{i}
=\frac{1}{V_0}\sum\limits_{\mathbf{q}}V_{i}(\mathbf{q})\rho(\mathbf{q}).
\end{equation}
$V_0$ is the volume of the system. The charge density operator is
defined as
$\rho(\mathbf{q})=\sum\limits_{\mathbf{k}}\psi^{\dagger}_{\mathbf{k}}\psi_{\mathbf{k+q}}$.
The electron-impurity scattering amplitude $V_{i}(\mathbf{q})$ could
be written as $v_{i}(\mathbf{q})\rho_{i}(-\mathbf{q})$, where
$\rho_{i}(-\mathbf{q})$ and $v_{i}(\mathbf{q})$ are the Fourier
components of the impurity density and the electron-impurity
potential, respectively. For charged impurity, $v_{i}(\mathbf{q})$
is taken as of the Thomas-Fermi type\cite{yan08,hwang09,yan09}
\begin{equation}
v_{i}(\mathbf{q})=\frac{2\pi e^{2}}{\epsilon(q+q_{TF})}e^{-qd}.
\end{equation}
$\epsilon$ is the effective dielectric constant from lattice and
substrate, $\epsilon$=3 is adopted in this
work\cite{hwang07b,yan08,yan09}. $d$ is the distance between the
impurities and the graphene plane and would be set as zero in the
present work\cite{hwang09,yan09}. $q_{TF}$ is the Thomas-Fermi
wave number and is obtained from the long-wavelength-limit static
polarizability of the corresponding noninteracting electron
system\cite{yan08,hwang09} as
\begin{equation}
q_{TF}=2\pi e^2\chi/\epsilon,
\end{equation}
with the static polarizability
\begin{equation}
\chi=\frac{2}{V_{0}}\int_{0}^{\beta}d\tau\langle\text{T}_{\tau}n(\tau)n^{\dagger}(0)\rangle_{c}.
\end{equation}
A factor of `2' comes from the two fold degeneracy in spin.
The subindex `$c$' means retaining only connected Feynman
diagrams in evaluating the expectation value. The particle
number operator is defined as
\begin{equation}
n(\tau)=\sum\limits_{\mathbf{k}}\psi_{\mathbf{k}}^{\dagger}(\tau)\psi_{\mathbf{k}}(\tau)
=\sum\limits_{\mathbf{k}}\varphi_{\mathbf{k}}^{\dagger}(\tau)\varphi_{\mathbf{k}}(\tau),
\end{equation}
where
$\varphi_{\mathbf{k}}^{\dagger}=(c_{1\mathbf{k}}^{\dagger},c_{2\mathbf{k}}^{\dagger},
c_{3\mathbf{k}}^{\dagger},c_{4\mathbf{k}}^{\dagger})$, with
$c_{\alpha\mathbf{k}}^{\dagger}$ representing the creation
operator of the $\alpha$-th ($\alpha$=1, 2, 3, 4) eigenstate of
$H_{0}(\mathbf{k})$ with eigenenergy denoted as
$\epsilon_{\mathbf{k}\alpha}$. Thus $\chi$ is obtained in terms
of the free particle eigenstates as
\begin{equation}
\chi=\frac{2\beta}{V_{0}}\sum\limits_{\mathbf{k}\alpha}n_{F}(\xi_{\mathbf{k}\alpha})n_{F}(-\xi_{\mathbf{k}\alpha}),
\end{equation}
where $\xi_{\mathbf{k}\alpha}$=$\epsilon_{\mathbf{k}\alpha}-\mu$
and $n_{F}(x)$=$1/(e^{\beta x}+1)$ is the Fermi distribution
function, $\mu$ is the chemical potential. $\beta$ represents
the inverse temperature $1/k_{B}T$, with $k_{B}$ the Boltzmann constant.
The value of $\chi$
and thus $q_{TF}$ depends on both the temperature and the
chemical potential.

In order to calculate the thermopower, we should first obtain
the particle current and heat current operators $\mathbf{j}_N$
and $\mathbf{j}_Q$. They are obtained in terms of the
continuity equation, which for the particle current
reads\cite{ambegaokar65,durst00}
\begin{equation}
\dot{\rho}(\mathbf{r})+\boldsymbol{\nabla}\cdot\mathbf{j}_{N}(\mathbf{r})=0.
\end{equation}
The momentum space version of the continuity function is
\begin{equation}
i\dot{\rho}(\mathbf{q})=[\rho(\mathbf{q}),\hat{H}]=\mathbf{q}\cdot\mathbf{j}_{N}(\mathbf{q}).
\end{equation}
A similar relationship holds for the energy density
$h_{E}(\mathbf{q})$ and the energy current
$\mathbf{j}_{E}(\mathbf{q})$. With the particle density
operator defined as
$\rho(\mathbf{q})=\sum\limits_{\mathbf{k}}\sum\limits_{\alpha=1}^{2}
(a^{\dagger}_{\alpha\mathbf{k}}a_{\alpha,\mathbf{k+q}}+
b^{\dagger}_{\alpha\mathbf{k}}b_{\alpha,\mathbf{k+q}})$, the
particle current operator is obtained at the $\mathbf{q}$=0
limit as
\begin{equation}
\mathbf{j}_{N}(\mathbf{q}=0)=\sum\limits_{\mathbf{k}}
\psi^{\dagger}_{\mathbf{k}}\mathbf{j}^{\mathbf{k}}_{1}\psi_{\mathbf{k}},
\end{equation}
where the matrix $\mathbf{j}^{\mathbf{k}}_{1}$ is defined as
\begin{equation}
\mathbf{j}^{\mathbf{k}}_{1}=\begin{pmatrix} 0 &
\mathbf{v}_{\mathbf{k}} & 0 & 0 \\ \mathbf{v}^{\ast}_{\mathbf{k}} &
0 & 0 & 0 \\ 0 & 0 & 0 & \mathbf{v}^{\ast}_{\mathbf{k}} \\ 0 & 0 &
\mathbf{v}_{\mathbf{k}} & 0 \end{pmatrix},
\end{equation}
where the velocity is defined as
$\mathbf{v}_{\mathbf{k}}$=$\boldsymbol{\nabla}_{\mathbf{k}}\phi(\mathbf{k})$.
The momentum space energy density operator could be written
as\cite{jonson80}
$h_E(\mathbf{q})=\sum\limits_{k}\psi^{\dagger}_{\mathbf{k}}h_{0}(\mathbf{k,\mathbf{q}})\psi_{\mathbf{k+q}}
+\frac{1}{V_{0}}\sum\limits_{\mathbf{q}'}V_{i}(\mathbf{q}')\rho(\mathbf{q'+q})$,
where the free part is
\begin{equation}
h_{0}(\mathbf{k},\mathbf{q})=\begin{pmatrix} \frac{V}{2} &
\phi(\mathbf{k},\mathbf{q}) & t_{\perp} & 0 \\
\phi^{\ast}(\mathbf{k},\mathbf{q}) &
\frac{V}{2} & 0 & 0 \\ t_{\perp} & 0 & -\frac{V}{2} &
\phi^{\ast}(\mathbf{k},\mathbf{q}) \\ 0 & 0
& \phi(\mathbf{k},\mathbf{q}) & -\frac{V}{2}
\end{pmatrix},
\end{equation}
where
$\phi(\mathbf{k},\mathbf{q})=(\phi(\mathbf{k})+\phi(\mathbf{k}+\mathbf{q}))/2$.
Calculating the commutator between $h_{E}(\mathbf{q})$ and the
Hamiltonian, the energy current operator could be obtained. The
heat current operator, defined by the $\mathbf{q}$=0 limit of
$\mathbf{j}_{Q}(\mathbf{q})=\mathbf{j}_{E}(\mathbf{q})-\mu
\mathbf{j}_{N}(\mathbf{q})$, is written as
\begin{equation}
\mathbf{j}_{Q}(\mathbf{q}=0)=\sum\limits_{\mathbf{k}}
\psi^{\dagger}_{\mathbf{k}}\mathbf{j}^{\mathbf{k}}_{2}\psi_{\mathbf{k}}
+\frac{1}{V_{0}}\sum\limits_{\mathbf{k},\mathbf{q}'}V_{i}(\mathbf{q}')\psi^{\dagger}_{\mathbf{k}}
\mathbf{j}^{\mathbf{k},\mathbf{q}'}_{1}\psi_{\mathbf{k+q'}},
\end{equation}
where the free part is written as
\begin{equation}
\mathbf{j}^{\mathbf{k}}_{2}=\begin{pmatrix} \mathbf{d}(\mathbf{k}) &
(\frac{V}{2}-\mu)\mathbf{v}_{\mathbf{k}} & 0 &
\frac{t_{\perp}}{2}\mathbf{v}^{\ast}_{\mathbf{k}} \\
(\frac{V}{2}-\mu)\mathbf{v}^{\ast}_{\mathbf{k}} &
\mathbf{d}(\mathbf{k}) &
\frac{t_{\perp}}{2}\mathbf{v}^{\ast}_{\mathbf{k}} & 0 \\ 0 &
\frac{t_{\perp}}{2}\mathbf{v}_{\mathbf{k}} & \mathbf{d}(\mathbf{k})
& -(\frac{V}{2}+\mu)\mathbf{v}^{\ast}_{\mathbf{k}} \\
\frac{t_{\perp}}{2}\mathbf{v}_{\mathbf{k}} & 0 &
-(\frac{V}{2}+\mu)\mathbf{v}_{\mathbf{k}} & \mathbf{d}_{\mathbf{k}},
\end{pmatrix}
\end{equation}
with
$\mathbf{d}(\mathbf{k})=\frac{1}{2}[\mathbf{v}_{\mathbf{k}}\phi^{\ast}(\mathbf{k})
+\mathbf{v}^{\ast}_{\mathbf{k}}\phi(\mathbf{k})]$. Substituting
$\mathbf{v}_{\mathbf{k}}$ by
$\frac{1}{2}(\mathbf{v}_{\mathbf{k+q'}}+\mathbf{v}_{\mathbf{k}})$,
we obtain $\mathbf{j}^{\mathbf{k,q'}}_{1}$ from
$\mathbf{j}^{\mathbf{k}}_{1}$.

We calculate the thermopower in terms of the Kubo's
formula\cite{jonson80,mahanbook} with impurity scattering treated to
the order of self-consistent Born
approximation\cite{lee93,shon98,koshino06,yan09}. The thermopower is
given by\cite{jonson80,callenbook,maekawabook}
\begin{equation}
S=-\frac{L_{12}}{eTL_{11}},
\end{equation}
where $e$ is the absolute value of the electron charge. The
linear response coefficients $L_{ij}$ are obtained from the
correlation function $\mathcal{L}_{ij}(i\omega)$ by
\begin{equation}
\mathit{L}_{ij}=\lim_{\omega\rightarrow
0}\text{Re}\mathcal{L}_{ij}(\omega+i0^{+}).
\end{equation}
In the Matsubara notation, the correlation function
reads\cite{jonson80}
\begin{equation}
\mathcal{L}_{ij}(i\omega_{n})=-\frac{iT}{(i\omega_{n})dV_{0}}
\int_{0}^{\beta} d\tau e^{i\omega_{n}\tau}\langle
T_{\tau}\mathbf{j}_{i}(\tau)\cdot\mathbf{j}_{j}(0)\rangle,
\end{equation}
where $d=2$ is the dimensionality, $\beta=1/k_{B}T$, and $T_{\tau}$
indicates an ordering of the current operators with respect to the
complex time $\tau$. $\omega_{n}$ is the bosonic Matsubara frequency
related with the current operator.

For charged impurities, only intravalley scattering is
important\cite{shon98,ostrovsky06,yan08,yan09}, so we neglect
the intervalley scattering processes in this work. Remembering
that only states close to the chemical potential contribute to
the \emph{dc} transport, we could focus our attention at a
single valley and retain only those low energy states. Here, we
would focus on the valley around
$\mathbf{K}=(\sqrt{3},1)\frac{2\pi}{3\sqrt{3}a}$. Similar to
the monolayer graphene case, this could be achieved by
introducing an energy cutoff $E_{C}$, such that only when the
smaller positive eigenenergy is less than $E_{C}$ would the
states labeled by the corresponding wave vector $\mathbf{k}$ be
retained in our calculations\cite{shon98,yan09}. For
experimentally relevant carrier densities, $E_{C}$ is no larger
than 1 eV. In this energy range, $\phi(\mathbf{k}')$ could be
expanded as a polynomial series in terms of the relative wave
vector $\mathbf{k}=\mathbf{k}'-\mathbf{K}$. Here, we retain the
expansion to the second order of $\mathbf{k}$ as
\begin{eqnarray}
\phi(\mathbf{k})&\equiv&\phi(\mathbf{k+K})=-t\sum\limits_{j=1}^3
e^{i(\mathbf{k+K})\cdot\boldsymbol{\delta}_{j}} \notag \\
&\simeq&\frac{3t}{2}e^{\frac{\pi}{3}i}[k_{y}a-ik_{x}a+\frac{1}{4}(k_{y}a+ik_{x}a)^{2}]
\end{eqnarray}
Test calculations show that the exact dispersion around
$\mathbf{K}$ could be excellently approximated by the above
approximation up to $E_{C}\approx 1.2$ eV, so it is accurate
enough for our problem.

In terms of the approximate $\phi(\mathbf{k})$, the velocity
vector satisfies
$\mathbf{q}\cdot(\mathbf{v_{k}}+\mathbf{v_{k+q}})/2
=\phi(\mathbf{k+q})-\phi(\mathbf{k})$. The impurity scattering
part of the heat current operator could thus be written as
$\frac{1}{V_{0}}\sum\limits_{\mathbf{q}}V_{i}(\mathbf{q})\mathbf{j}_{N}(\mathbf{q})$.

Since the particle current operator is the same as that without
of impurity scattering, the linear response coefficient
$L_{11}$ could easily be shown to
be\cite{jonson80,shon98,yan08,yan09}
\begin{eqnarray} \label{L11}
L_{11}=& &T\int_{-\infty}^{+\infty}\frac{d\epsilon}{2\pi}
[-\frac{\partial n_{F}(\epsilon)}{\partial \epsilon}]\text{Re}\{P_{11}(\epsilon-i0^{+},\epsilon+i0^{+}) \notag\ \\
& & -P_{11}(\epsilon+i0^{+},\epsilon+i0^{+})\}.
\end{eqnarray}
The kernel is defined as
\begin{equation}
P_{11}(z,z')=\frac{2}{d V_{0}}\sum\limits_{\mathbf{k}}\text{Tr}
\{G_{\mathbf{k}}(z)\boldsymbol{\Gamma}_{1}(\mathbf{k},z,z')G_{\mathbf{k}}(z')\cdot\mathbf{j}_{1}^{\mathbf{k}}\},
\end{equation}
with $\boldsymbol{\Gamma}_{1}(\mathbf{k})$ as the vertex function corresponding to the wave vector $\mathbf{k}$.

Taking into account of the relationship
$\mathbf{j}_{1}^{\mathbf{k,q}}
=\frac{1}{2}\mathbf{j}_{1}^{\mathbf{k}}+\frac{1}{2}\mathbf{j}_{1}^{\mathbf{k+q}}$,
and following the same route as for the single orbital
model\cite{jonson80}, it could be proved that $L_{12}$ could be
written into the following form
\cite{jonson80,yan09,lofwander07}
\begin{eqnarray}
L_{12}=&&T\int_{-\infty}^{+\infty}\frac{d\epsilon}{2\pi}
[-\frac{\partial n_{F}(\epsilon)}{\partial\epsilon}] \text{Re}\{P_{12}(\epsilon-i0^{+},\epsilon+i0^{+}) \notag \\
&&-P_{12}(\epsilon+i0^{+},\epsilon+i0^{+})\}.
\end{eqnarray}
The kernel $P_{12}$ is simply $P_{11}$ multiplied by the energy
\begin{equation}
P_{12}(\epsilon\mp i0^{+},\epsilon+i0^{+})=\epsilon P_{11}(\epsilon\mp i0^{+},\epsilon+i0^{+}).
\end{equation}
In our calculations, the positive infinitesimal $0^{+}$ would
be replaced by a small positive quantity $\eta$. In this work,
we adopt $\eta$=1 meV. A smaller $\eta$ is verified not to
change the results to be presented in what follows.

Under the SCBA, the $4\times 4$ Green's function matrix is
defined as
$G_{\mathbf{k}}(z)=(G^{0}_{\mathbf{k}}(z)^{-1}-\Sigma_{\mathbf{k}}(z))$.
The self energy is determined by the following self-consistency
relation
\begin{equation}
\Sigma_{\mathbf{k}}(z)=\frac{n_{i}}{V_{0}}\sum\limits_{\mathbf{k}'}|v_{i}(\mathbf{k-k'})|^{2}
[G_{\mathbf{k}'}^{0}(z) -\Sigma_{\mathbf{k}'}(z)]^{-1},
\end{equation}
where $G_{\mathbf{k}}^{0}(z)=[z+\mu-H_{0}(\mathbf{k})]^{-1}$ is
the free Green's function. $\mu$ is the chemical potential
determined from the free Hamiltonian for a certain carrier
density. In this work, we would neglect the shift in $\mu$ by
the impurity potential. This is known as not giving rise to
qualitative changes when the impurity concentration is not very
high\cite{yan09}. After obtaining the Green's functions, the
vertex functions are calculated by the following
self-consistency relation
\begin{equation}
\boldsymbol{\Gamma}_{1}(\mathbf{k},z,z')=\mathbf{j}_{1}^{\mathbf{k}}+\frac{n_{i}}{V_{0}}
\sum\limits_{\mathbf{k'}}|v_{i}(\mathbf{k-k'})|^{2}G_{\mathbf{k}'}(z)\boldsymbol{\Gamma}_{1}(\mathbf{k'},z,z') G_{\mathbf{k}'}(z').
\end{equation}
In obtaining the above self-consistent relations, averages over
impurity configurations have been done under SCBA
as\cite{mahanbook}
\begin{equation}
\langle\rho_{i}(\mathbf{q})\rho_{i}(-\mathbf{q}')\rangle=N_{i}\delta_{\mathbf{q,q'}},
\end{equation}
where $N_{i}=n_{i}V_{0}$ is number of impurities in the system
under consideration. For a set of $\mathbf{k}$ vectors, we can
get the Green's functions and the vertex functions, then they
are used to calculate the kernels $P_{11}$ and $P_{12}$, and
hence the linear response coefficients $L_{11}$ and $L_{12}$
could be obtained. If the full Brillouin zone (BZ) is utilized,
the number of wave vectors would be too large for a practical
calculation. However, since only intravalley scattering is
relevant for charged
impurities\cite{nomura07,hwang07,adam07,yan09}, we could focus
on the low energy states within a cutoff energy $E_{C}$ around
a single valley.

\section{result and discussion}
To see the effect of impurity scattering more clearly, we first
present the results for clean systems, where the bare Green's
functions and vertex functions are used. The room temperature
(300 K) thermopower of clean monolayer and gapless bilayer
graphene are presented in Fig 1(a). Here and later, the
abscissa index $x$ represents the electron doping averaged to
per site, which could be controlled by an external gate
voltage\cite{zuev09,wei09,checkelsky09}. $x$=0.001 amounts to
an electron density of 3.82$\times$10$^{12}$ cm$^{-2}$ per
layer. For the monolayer system\cite{ouyang09}, a tight binding
model up to nearest-neighbor hopping $t$=3 eV is used. Peak
values of the two curves are almost the same, with that of the
monolayer slightly larger. When a more realistic parameter
$t$=2.7 eV is used for the monolayer graphene\cite{peres06},
the peak position shifts slightly away from $x$=0, while the
peak value keeps almost unchanged as about $\pm$83 $\mu$V/K.
Besides the low carrier density peak, thermopower of gapless
bilayer graphene shows a second smaller peak at a higher
carrier concentration. The onset of the second peak, for which
the peak position corresponds to a chemical potential of
$|\mu|$$\simeq$316 meV, is associated with the crossing of the
chemical potential with the lower valence band (band top at
-$t_{\perp}$=-300 meV) or the upper conduction band (band
bottom at $t_{\perp}$=300 meV).

\begin{figure}
\centering
\includegraphics[width=10cm,height=15cm,angle=270]{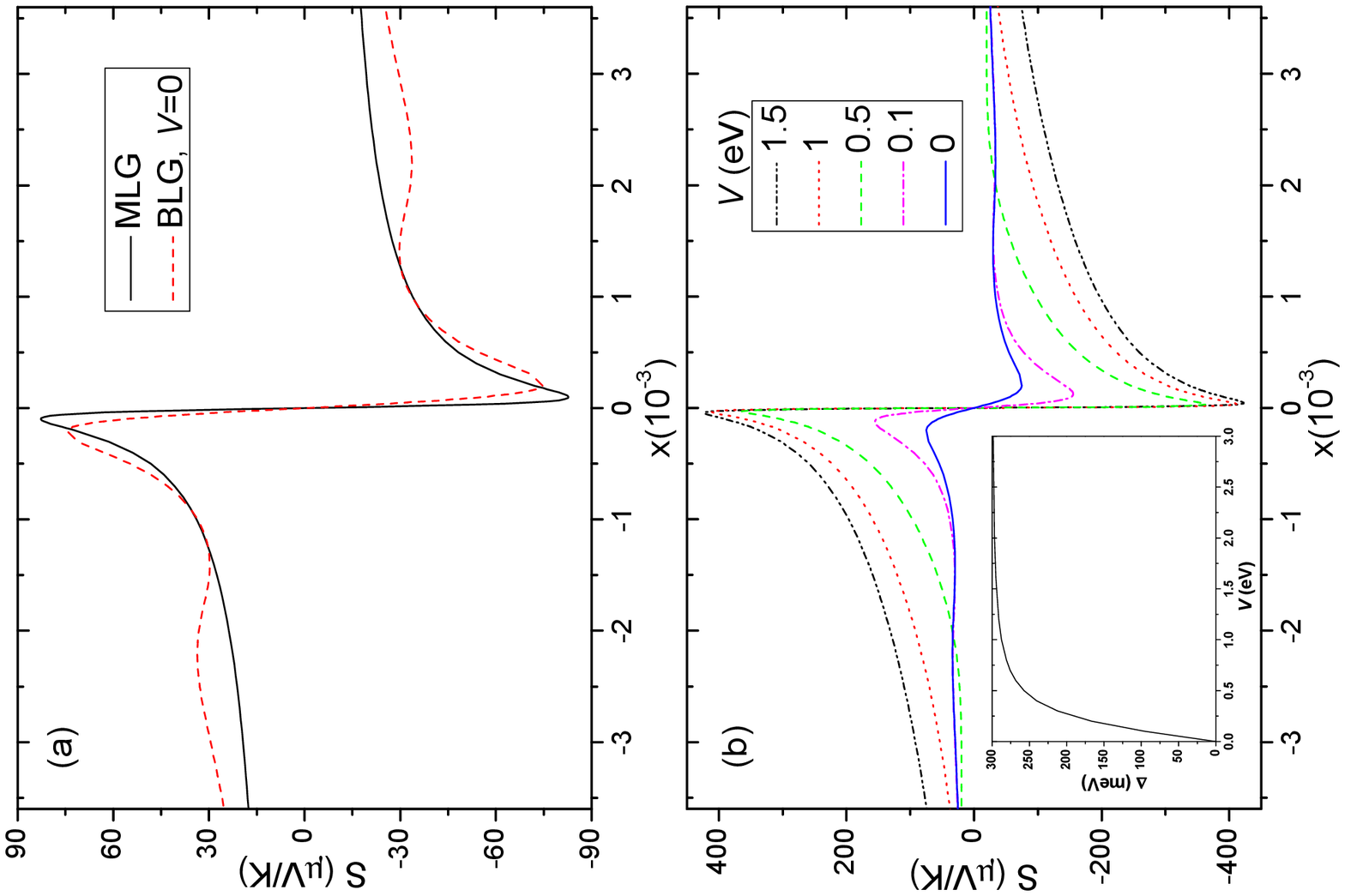}
\caption{(a) Room temperature thermopower as a function of carrier density for clean
monolayer graphene (MLG) and gapless bilayer graphene (BLG).
(b) Room temperature thermopower of BLG for a series of potential
energy differences $V$ between the two layers. Inset of (b) shows
the evolution of the global energy gap $\Delta$ as a function of $V$.}
\end{figure}

We now explore the effect of opening a band gap in the clean
bilayer graphene. Figure 1(b) shows the thermopower of bilayer
graphene for a series of potential differences $V$ between the
two layers. With the increase of $V$, peak value of the
thermopower increases quickly. For $V=1$ eV (corresponding to
an energy gap of approximately 288 meV, which is experimentally
achievable\cite{zhang09}), peak value of the thermopower is
about 412 $\mu$V/K, which is more than four times that of the
value in monolayer graphene and zero gap bilayer graphene. The
much smaller second peak shifts continuously to larger $x$ as
$V$ increases and becomes irrelevant as $V$ increases to 1 eV.
Hence, we would concentrate on the region of low carrier
density in the following. The smallest energy gap between the
conduction band and the valence band increases with $V$ as
$\Delta=\frac{Vt_{\perp}}{\sqrt{t_{\perp}^2+V^2}}$
\cite{nilsson07,dahal08}, which is shown as an inset of Fig.
1(b). The large magnitude of thermopower and the tunability of
gap make the biased bilayer graphene a more promising candidate
for future thermoelectric applications as compared with
monolayer graphene.

In the following, we focus on the biased bilayer graphene
system with $V$=1 eV. In Fig. 2(a), thermopower of this system
is shown as a function of chemical potential ($\mu$) for three
different temperatures. When $\mu$ is at the band edge ($\sim$
$\pm$0.144 eV), thermopower of all three temperatures are
nearly identical to each other. When $\mu$ is inside of the
band gap, $|S|$ increases as temperature decreases. While when
$\mu$ lies in the bands, $|S|$ increases as temperature
increases. This is similar to the corresponding behavior in
semiconducting armchair graphene
nanoribbons\cite{ouyang09,xing09}. The corresponding dependence
of $x$ on $\mu$ is illustrated in Fig. 2(b). When $x$ is very
close to zero, $|\mu|$ decreases as temperature increases. As
$|x|$ increases beyond a certain critical value, $|\mu|$
increases as temperature increases. The above complex
temperature dependence is a direct result of the existence of
the band edge Van Hove singularities. From Fig. 2(a) and Fig.
2(b), it is clear that for 100 K, peak value of $S$ is achieved
at a doping very close to zero. As temperature increases, the
carrier density for the peak increases continuously to larger
values.

\begin{figure}
\centering
\includegraphics[width=10cm,height=15cm,angle=270]{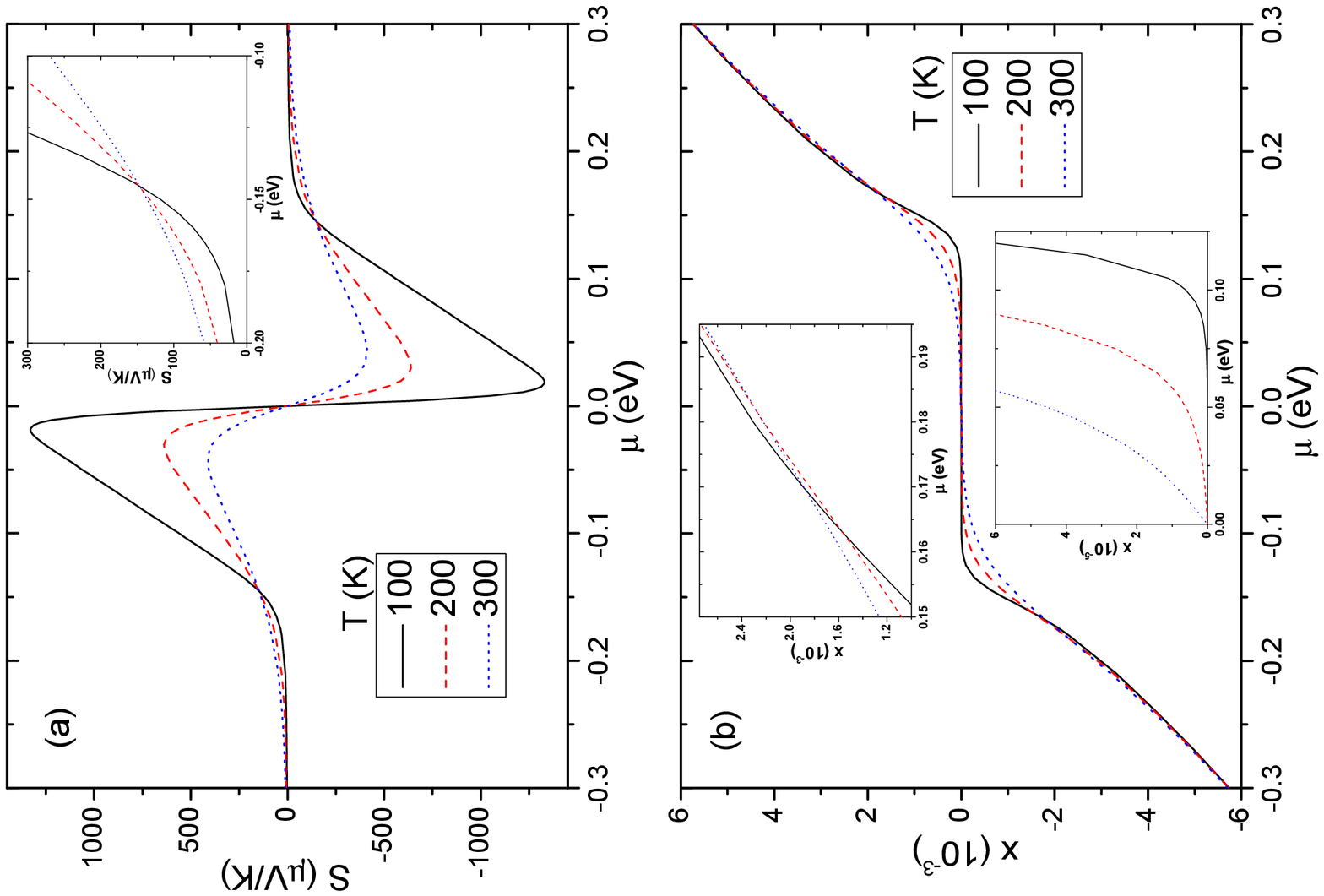}
\caption{(a) Thermopower as a function of chemical potential for a
biased bilayer graphene, at three different temperatures.
(b) The corresponding variation of doping concentration $x$ as a
function of the chemical potential. Inset of (a) shows an enlargement
of the curves close to the valence band top ($\sim$-0.144 eV). Lower inset of (b) shows
an enlargement of the small $\mu$ and small $x$ region. Upper inset
of (b) shows an enlargement of the region where the
temperature dependence of $\mu$ for fixed $x$ changes.}
\end{figure}

Now, we begin to study the effect of impurity. In this work,
charged impurity is considered as the only source of scattering
for impure bilayer graphene. Thermopower of bilayer graphene
for a series of different impurity concentrations are shown in
Fig. 3(a), at room temperature for $V=1$ eV. Curve for clean
system is also displayed as a reference. In order to ensure
that the SCBA is valid, we consider only cases with small
impurity concentrations. Up to a concentration of
$n_i$=5$\times$10$^{10}$ cm$^{-2}$ per layer, peak value of the
thermopower increases continuously with the impurity
concentration. The peak position remains unchanged. In these
relatively clean systems, the influence of
localization\cite{nilsson07,koshino08b,mkhitaryan08} can be
safely neglected.

\begin{figure}
\centering
\includegraphics[width=10cm,height=15cm,angle=270]{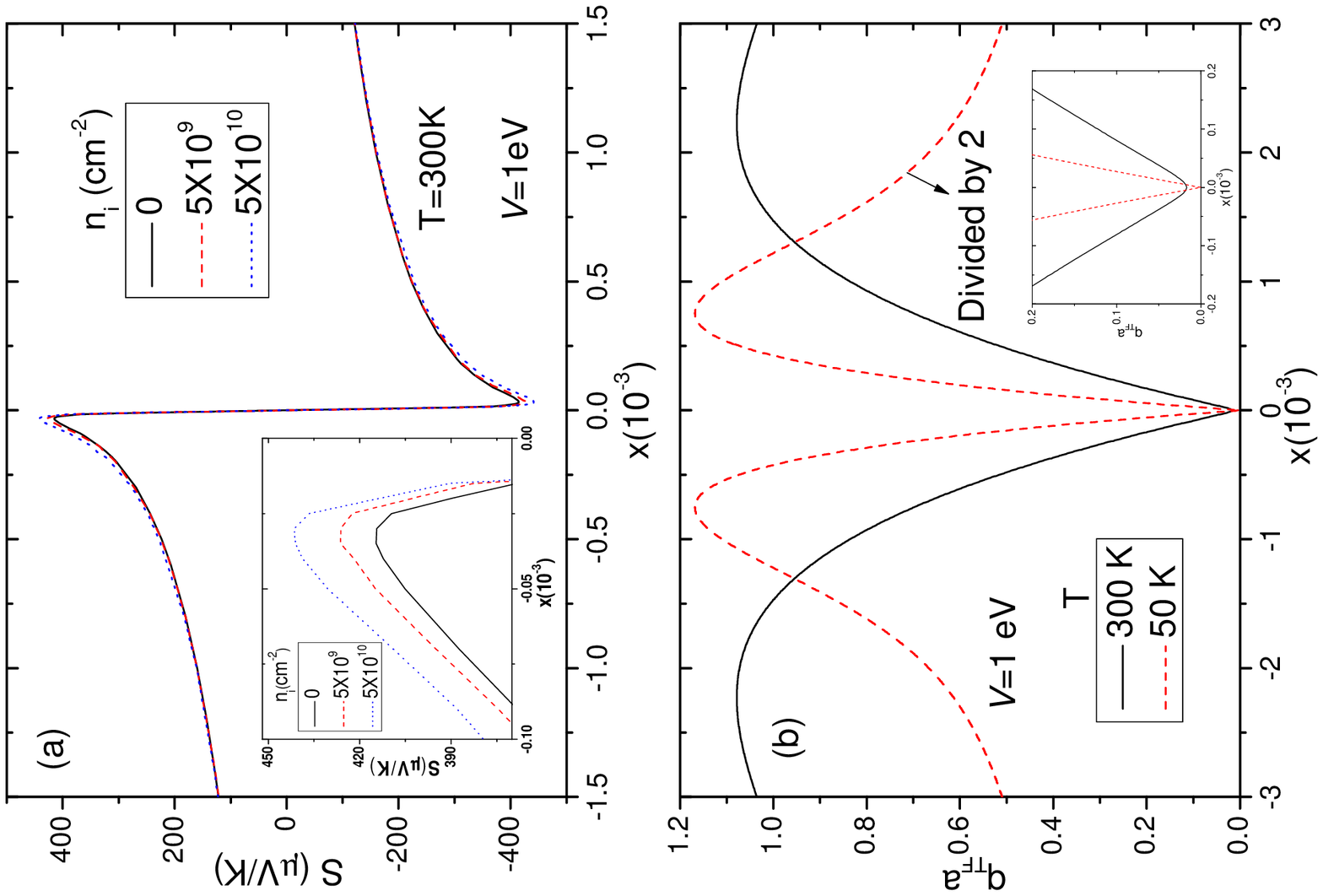}
\caption{(a) Room temperature thermopower of bilayer graphene
for a series of different impurity concentrations with
$V$=1 eV. (b) The Thomas-Fermi screening wave vector at 300 K and 50 K
for bilayer graphene with $V$=1 eV. The curve for 50 K is divided by a
factor of 2. Insets of (a) and (b) show an enlargement
of the corresponding low carrier density parts.}
\end{figure}

Here and later when impurity scattering is considered, an
energy cutoff of $E_{C}$=0.5 eV is used. Test calculations by
increasing $E_{C}$ show no perceivable change in the results
for both clean and impure systems in the considered low carrier
density region. For most energies, the self-consistency for the
Green's functions and vertex functions converge within 100
iterations with an accuracy of $10^{-5}$ and $10^{-4}$ for the
modulus of every element, respectively. For a symmetric band
structure, as is the case for both clean monolayer and bilayer
graphene, thermopower is an odd function of the carrier
density\cite{hwang09,yan09}. In the presence of charged
impurities, the electron-hole symmetry of the band is preserved
which could be seen by explicitly calculating the density of
states. So we expect the relationship $S(-x)=-S(x)$ survives.
The full curve for $n_{i}$=5$\times$10$^{10}$ cm$^{-2}$ and
$T=300$ K is calculated explicitly, verifying the above
statement. To save computing time, for all other parameter sets
in the presence of impurity, thermopower is calculated
explicitly for the hole doped cases with $x\le0$. The results
for $x>0$ are obtained through $S(x)=-S(-x)$.

The Thomas-Fermi screening wave vector is shown in Fig. 3(b) as
a function of carrier density at 300 K and 50 K for $V$=1 eV.
At 300 K, as a result of thermal excitations, we get a finite
$q_{TF}a$$\simeq$0.017 for zero doping. As temperature goes
down, the zero doping Thomas-Fermi screening wave vector
decreases gradually. As could be seen in the inset of Fig.
3(b), $q_{TF}(x=0)$ is very close to zero at 50 K. This result
is similar to the monolayer graphene system and is different
from the gapless bilayer graphene, for which the zero doping
$q_{TF}$ is finite even at zero temperature for the
nonvanishing density of states there\cite{sarma09}. The two
peaks in Fig. 3(b) arise from the Van Hove singularities near
the conduction band bottom and the valence band top, and are
shifted to larger carrier densities at higher temperature by
thermal excitations.

Previous
works\cite{ouyang09,yan09,hwang09,zuev09,wei09,checkelsky09} on
monolayer graphene show that impurity scattering is essential
to reproduce the temperature dependence of the thermopower
observed experimentally. Thermopower of gapped bilayer graphene
with the same impurity concentration $n_{i}$=5$\times$10$^{10}$
cm$^{-2}$ are shown as a function of carrier density for room
temperature and 50 K in Fig. 4. Results for $x\ge$10$^{-5}$ are
readily obtained for both temperatures. However, for $x$ at and
very close to zero, only results for $T$=300 K are obtained
within our calculation time. This is understood from Fig. 3(b)
as a result of reduced screening at a temperature as low as 50
K\cite{yan09}. As temperature decreases, thermopower is
suppressed and peak position of thermopower shifts slightly
towards zero doping. These results are qualitatively very
similar to the experimental results for monolayer
graphene\cite{zuev09,wei09,checkelsky09}.

\begin{figure}
\centering
\includegraphics[width=8cm,height=11cm,angle=270]{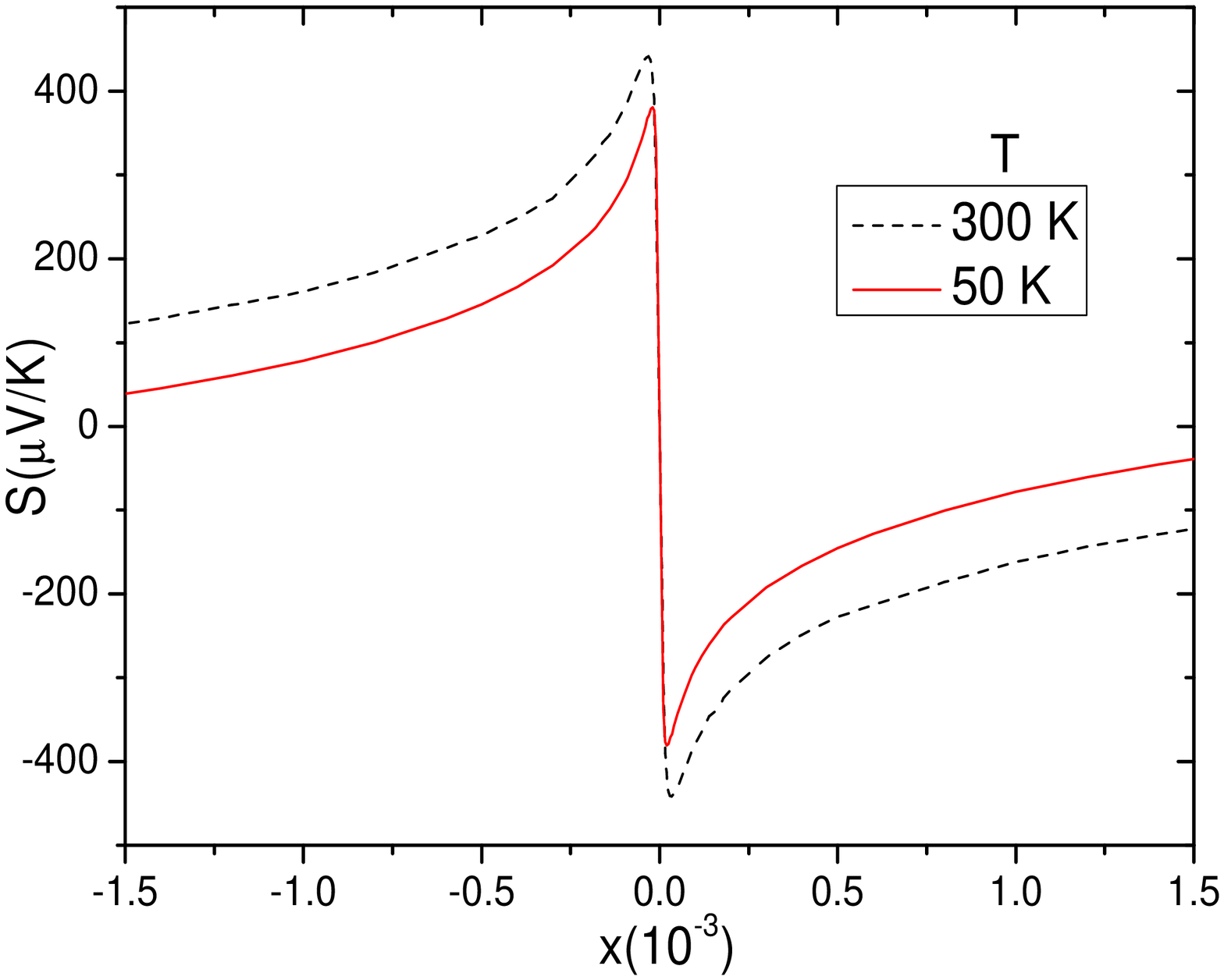}
\caption{Thermopower of gapped bilayer graphene at 300 K and 50 K
as a function of carrier density, for $V$=1 eV and $n_{i}$=5$\times$10$^{10}$ cm$^{-2}$.}
\end{figure}

A peculiar feature of our results is that the maximum value of
$S$ at room temperature increases with impurity concentration
$n_{i}$, as shown in Fig. 3(a). By increasing the energy cutoff
$E_{C}$ and the number of wave vectors in the Brillouin zone,
we have verified that the above result is robust. In order to
have a better understanding, we show the variation of $L_{11}$
and $L_{12}$ for $x$=-3$\times$10$^{-5}$ in Fig. 5(a). It is
clear that, both $L_{11}$ and $L_{12}$ decrease sharply as
$n_{i}$ increases. However, the reduction of $L_{11}$ is
somewhat larger than that of $L_{12}$. The increase of room
temperature $S$ thus comes as a result of the stronger
dependence of $L_{11}$ on $n_{i}$ as compared with $L_{12}$.
The reason why $L_{11}$ decreases faster than $L_{12}$ as
$n_{i}$ increases is encoded in the integration kernels of the
two linear response coefficients, Eq. (21) and Eq. (23).  Eq.
(23) shows that states above and below of the chemical
potential contribute to $L_{12}$ in opposite sign but of the
same sign for $L_{11}$. Hence $L_{11}$ is more sensitive to the
variation of density of states (DOS)  in the conduction band
and valence band due to the presence of impurities which is
shown in Fig. 5(b) for two impurity concentrations at $V$=1 eV
and $T$=300 K. In a previous theoretical work on graphene
nanoribbon\cite{ouyang09}, a similar increase of $S$ with
defect density is also observed.

\begin{figure}
\centering
\includegraphics[width=10cm,height=15cm,angle=270]{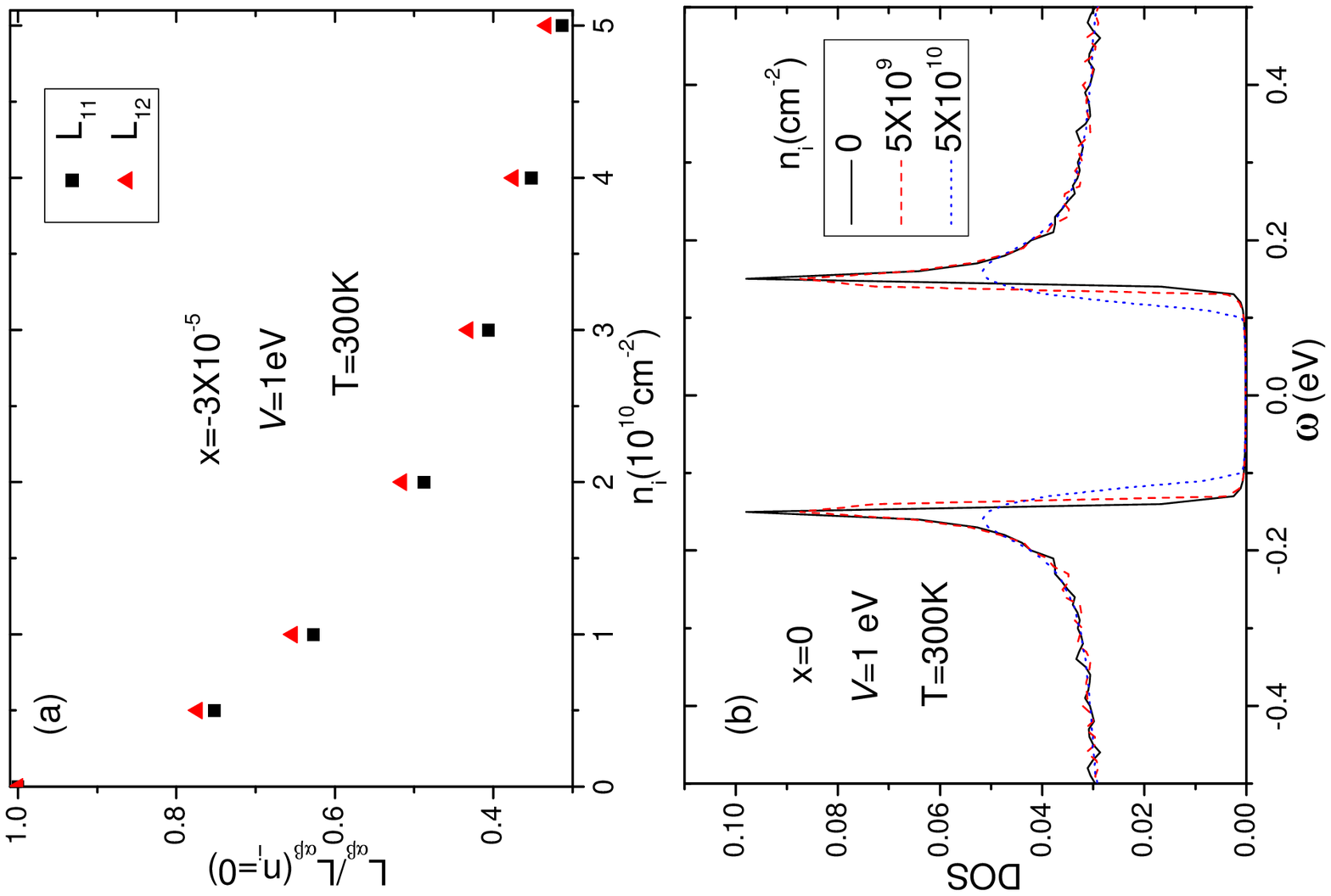}
\caption{(a) Variation of $L_{11}$ and $L_{12}$ as a function of
impurity concentration for $x$=-3$\times$10$^{-5}$,
normalized by the $n_{i}$=0 value, respectively.
(b) Low energy density of states (DOS) for three different impurity concentrations.}
\end{figure}

We now try to understand the doping dependence of the shape of
thermopower in the small carrier density region. From the
definitions in Eqs. (16), (20) and (22), thermopower could be
understood as the average value of $(E-\mu)$ weighted by the
combination of electron group velocity (encoded in
$\mathbf{v_{k}}$ in the current operator and the renormalized
current vertex) and DOS \cite{cutler69,ouyang09}. States above
and below of the chemical potential contribute in the opposite
sign to $S$. At the same time, the factor $\partial
n_{F}/\partial\omega$=$-n_{F}(1-n_{F})/k_{B}T$ is substantially
nonzero only in an energy range of several $k_{B}T$ centered
symmetrically around the chemical potential. It is thus easy to
understand that $S$ tends to be larger once the DOS and (or)
group velocity between states above and below the chemical
potential have a big contrast\cite{kuroki07}. According to this
picture, when $\mu$ is deep inside of the band (but still in
the low carrier density region), where the DOS is nearly flat
according to Fig. 5(b), the thermopower would be small. As the
chemical potential moves close to the band edge, difference
between states above and below of the chemical potential
increases, which results in the increase of $S$. At finite
temperature, when the doping is at or very close to the band
edge Van Hove singularities, $\mu$ would be inside of the gap.
In these cases, all states are high energy states measured from
$\mu$. Since $S$ would be larger once the higher $|E-\mu|$
states contribute more to the integration, $S$ is expected to
continue increase for very low carrier densities. As shown in
previous works\cite{cutler69,ouyang09}, for $\mu$ inside of the
semiconducting gap, the most significant part of $S$ comes from
a term of the form
\begin{equation}
S\sim(\frac{\Delta}{2}-|\mu|)/eT,
\end{equation}
which clearly shows the increase of $S$ as $\mu$$\rightarrow$0
(or equivalently, $x$$\rightarrow$0).

The above picture explains the initial increase of thermopower
as $|x|$ decreases. However, when the carrier density is very
close to zero, the chemical potential also lies close to zero
energy. Hence, both the valence band and the conduction band
states are present by thermal excitations. Since the
contributions to thermopower from valence and conduction band
states are opposite in sign, the thermopower is expected to
decrease at a critical carrier density characterized by the
temperature\cite{zuev09,wei09,ouyang09}. The critical $|x|$
below which the magnitude of thermopower starts to decrease is
thus expected to increase with temperature. This explains the
shift of peak position as observed in Fig. 2 and Fig. 4. In
Ref. 7 on thermopower of monolayer graphene, the deviation at
low carrier density from the higher density Mott's behavior
\cite{cutler69,hwang09} is ascribed to impurity scattering
mediated coherence between the conduction and valence bands.
According to this mechanism, as impurity concentration $n_{i}$
increases, the above coherence effect should enhance. So a
shift of peak position with $n_{i}$ is expected. At 50 K and
for $n_{i}$=5$\times$10$^{10}$ cm$^{-2}$, the peak position is
shown to shift from $x_{C}$$\simeq$0 to
$x_{C}$$\simeq$2$\times$10$^{-5}$. However, for 300 K, the
$x_{C}$ show no perceivable variation with $n_{i}$ up to
$n_{i}$=5$\times$10$^{10}$ cm$^{-2}$. A calculation beyond the
SCBA is needed to know whether or not the peak position for 300
K would shift for much larger $n_{i}$.

It is also interesting to ask why introducing a gap
significantly enhances the thermopower of bilayer graphene.
Formerly, a `pudding mold' mechanism\cite{kuroki07} is
introduced to account for the large thermopower observed in
Cobaltates. In that model, a band with a somewhat flat portion
connected to a highly dispersive portion is proposed to give
high thermopower when the chemical potential lies close to the
bending point.
Band structure of biased bilayer
graphene is exactly of `pudding mold'
like\cite{mccann06,kuroki07}. So the increase of $S$ with $V$
for a carrier density typically of $x$=$\pm$0.001, for which
the chemical potential lies inside of the band, is understood
as resulting from the onset of the `pudding mold' mechanism.
However, peak value of $S$ occurs when $\mu$ situates inside of
the band gap. In this case, as mentioned above, $S_{max}$ could
be estimated by Eq. (27)\cite{cutler69,ouyang09}. For $V$=1 eV,
$\Delta$$\simeq$288 meV. For $x$=-3$\times$10$^{-5}$ and
$T$=300 K, the chemical potential $\mu$$\simeq$-40 meV, Eq.
(27) gives a value of approximately 347 $\mu$V/K which is about
80 percent of the values in Fig. 2(a). So in the present case,
$S_{max}$ is set by the energy gap $\Delta$, which increases
with $V$ and is bounded by a limit,
$t_{\perp}$\cite{nilsson07,dahal08}. Hence, the large maximum
thermopower in biased bilayer graphene is mainly a result of
the energy gap.

Taking into account of the band asymmetry arising from the
on-site energy difference between the two kinds of carbon
sublattices, the band gap becomes asymmetric and the conduction
(or the valence, depending on the sign of $V$) band would be
more flat\cite{li09}. However, since the on-site energy
difference derived from experiment is only about 0.018
eV\cite{li09}, we expect the above effect is extremely small
and would not change our present result much. The interlayer
hopping is estimated to be in the range of
$t_{\perp}$$\sim$$0.3-0.4$ eV \cite{mcclure57}. Our test
calculation in clean system with $t_{\perp}$=0.4 eV for $V$=1
eV (with other parameters unchanged) gives a peak thermopower
value of approximately 544 $\mu$V/K at 300 K, which is 132
$\mu$V/K larger than the result for $t_{\perp}$=0.3 eV. So, the
thermopower of gapped bilayer graphene is large regardless of
the choice of model parameters.

Applying an electric field to trilayer or other multilayer
graphene system, a gap could also be
induced\cite{avetisyan09,koshino09}. For certain parameters,
the band structure is also of `pudding mold'
like\cite{avetisyan09,koshino09}. It is thus interesting to ask
how the peak value of thermopower evolves as a function of the
layer number. In this work, we have only considered the effect
of charged impurity scatterings. Recently, it has been proposed
that scattering by short-range disorder may also play an
important role in the transport of gapless bilayer graphene
because the screening of charged impurities in zero gap bilayer
graphene is much stronger than that in monolayer
graphene\cite{xiao09,sarma09}. Though we believe that the
degree of screening in gapped bilayer graphene should be much
smaller than that in the gapless system, it is an interesting
question whether the inclusion of short-range scatterers would
change our present results much. On the other hand, the regime
of large impurity concentration or strong impurity strength
deserves an explicit study in terms of a less severe
approximation as compared with SCBA used above. We defer the
above questions to later studies.

\section{summary}
We have theoretically studied the thermopower of bilayer
graphene. If a band gap of approximately 288 meV is induced in
the system by an external bias, the room temperature
thermopower is greatly enhanced by a factor of larger than 4 as
compared with that of the monolayer graphene and the gapless
bilayer graphene. In the presence of dilute charged impurities,
peak value of the room temperature thermopower is shown to
increase slightly. This behavior is analyzed in terms of the
different dependence of $L_{11}$ and $L_{12}$ on the
modification of density of states by impurity. As temperature
decreases, peak position of thermopower shifts slightly towards
zero carrier density in the presence of dilute charged
impurities.

\begin{acknowledgments}
We wish to acknowledge the support of NSC
98-2112-M-001-017-MY3. Part of the calculations was performed
in the National Center for High-Performance Computing in
Taiwan.
\end{acknowledgments}\index{}


\end{document}